\documentstyle[psfig,pre,multicol,aps]{revtex}

\def\sech{\mbox{sech}} 
\def\ds{\displaystyle}
\def\un{\underline}

\begin{document}

\draft

\title{Models for energy and charge transport\\
and storage in biomolecules}

\author{S.F. Mingaleev}
\address{Bogolyubov Institute for Theoretical Physics, 
14-b Metrologichna Str., 252143 Kiev, Ukraine}
\author{P.L. Christiansen}
\address{Department of Mathematical Modelling, 
The Technical University of Denmark, 
DK-2800 Lyngby, Denmark}
\author{Yu.B. Gaididei}
\address{Bogolyubov Institute for Theoretical Physics, 
14-b Metrologichna Str., 252143 Kiev, Ukraine}
\author{M. Johansson}
\address{Department of Physics and Measurement 
Technology, Link{\"o}ping University, S-581 83 
Link{\"o}ping, Sweden}
\author{K.{\O}. Rasmussen}
\address{Theoretical Division, Los Alamos National 
Laboratory, Los Alamos, New Mexico 87545, USA}

\date{August 25, 1998}
\maketitle

\begin{abstract}
Two models for energy and charge transport and 
storage in biomolecules are considered. 
A model based on the discrete nonlinear Schr{\"o}dinger 
equation with long-range dispersive interactions (LRI's) between 
base pairs of DNA is offered for the description of nonlinear 
dynamics of the DNA molecule. We show that LRI's are responsible 
for the existence of an interval of bistability where two stable 
stationary states, a narrow, pinned state and a broad, mobile state, 
coexist at each value of the total energy. The possibility of controlled 
switching between pinned and mobile states is demonstrated. The 
mechanism could be important for controlling energy storage 
and transport in DNA molecules. 
Another model is offered for the description of nonlinear excitations in 
proteins and other anharmonic biomolecules. We show that in the {\em highly 
anharmonic} systems a bound state 
of Davydov and Boussinesq solitons can exist. 
\end{abstract}

\vspace{6mm}

\vspace{3mm}
\noindent
{\bf Published:} \ Journal of Biological Physics {\bf 25}, 41--63 (1999).


\begin{multicols}{2}
\narrowtext

\section{Introduction}

Understanding how biological macromolecules (proteins, DNA, RNA, etc.)
function in the living cells remains the major challenge in molecular 
biology. 
One of the most important questions is the mechanism of gene
expression. The expression of a given gene involves two steps: 
transcription and translation. The transcription includes copying the 
linear genetic information into the messenger ribonucleic acid (mRNA). 
The information stored in mRNA is transferred into a sequence of aminoacids 
using the genetic code. mRNA is produced by the enzyme RNA-polymerase (RNAP) 
which binds to the promoter segment of DNA. As a result of the interaction
between RNAP and promoter of DNA the so-called "bubble" (i.e. a state in 
which 10--20 base pairs are disrupted) is formed. The disruption of 
20 base pairs corresponds to investing some 100 kcal/mole (0.43 eV) 
\cite{reiss}. 

In the framework of a linear model the large-amplitude motion of the bases 
was supposed to occur due to an interference mechanism \cite{loz79}. 
According to this model energetic solvent molecules kick DNA and create 
elastic waves therein. As a result of the interference of two counter 
propagating elastic waves, the base displacements may exceed the elasticity 
threshold  such that DNA undergoes a transition to a kink form which is more 
flexible. A similar approach was also proposed in Refs. \cite{cm88,pr86}. The 
linear elastic waves in DNA are assumed to be strong enough to break a 
hydrogen bond and thereby facilitate the disruption of base pairs. In spite 
of the attractiveness of this theory, which gives at least a qualitative 
interpretation of the experimental data \cite{geor96}, there are the following 
fundamental difficulties which to our opinion are inherent in the linear 
model of the DNA dynamics:
 (i) The dispersive properties (the dependence of the group velocity on the 
wave length) of the vibrational degrees of freedom in DNA will cause 
spreading of the wave packets and therefore smear the interference pattern. 
Furthermore, it has been shown \cite{hk84} that the amplitudes of the sugar 
and the base vibrations are rather large even in a crystalline phase of 
DNA. Since the large-amplitude vibrations in the molecules and the molecular 
complexes are usually highly anharmonic their nonlinear properties can not 
be ignored; 
  (ii) Molecules and ions which exist in the solution permanently interact 
with DNA. These interactions are usually considered as white noise and their 
influence is modelled by introducing Langevin stochastic forces  into the 
equations describing the intramolecular motion. It is well known \cite{gs76} 
that stochastic forces provide relaxation of linear excitations and 
destroy their coherent properties. Equivalently the coherence length (the 
length of the concerted motions) rapidly decreases with increasing 
temperature;
  (iii) DNA is a complex system which has many nearly isoenergetic 
ground states  and may therefore be considered as a fluctuating 
aperiodic system. DNA may have physical characteristics in common with 
quasi-one-dimensional disordered crystals or glasses. However, it is 
known \cite{pap71} that the transmission coefficient for a linear wave 
propagating in a disordered chain decreases exponentially with the 
growth of the distance (Anderson localization). In this way it is 
difficult to explain in the framework of linear theory such a phenomenon 
as an action at distance, where concerted motion initiated at one end of a 
biological molecule can be transmitted to its other end. 

The above mentioned fundamental problems can be overcome in the framework 
of nonlinear models of DNA. 
Nonlinear interactions can give rise to very stable excitations, called 
solitons, which can travel without changing their shape. These excitations 
are very robust and important in the coherent transfer of energy 
\cite{wa76}.  For realistic interatomic potentials the solitary waves 
are compressive and supersonic. They propagate without energy loss, and 
their collisions are almost elastic. 

Nonlinear interactions between atoms in DNA can give rise to intrinsically 
localized breather-like vibration modes \cite{st88,au94}. Such localized 
modes,  
being large-amplitude vibrations of a few (2 or 3) particles, can 
facilitate the disruption of base pairs and in this way initiate 
conformational transitions in DNA. These modes can occur as a result 
of modulational instability of continuum-like nonlinear modes 
\cite{pou93}, which is  created by energy exchange mechanisms between the 
nonlinear excitations. The latter favors the growth of the large 
excitations \cite{dp93}.

Nonlinear solitary excitations can maintain their overall shape on long 
time scales even in the presence of the thermal fluctuations. Their robust 
character under the influence of white noise was demonstrated \cite{mu90} 
and a simplified model of double-stranded DNA was proposed and explored. 
Quite recently the stability of highly localized, breather-like, 
excitations in discrete nonlinear lattices under the influence of thermal 
fluctuations was investigated \cite{ch96}. It was shown that the lifetime 
of a breather increases with increasing nonlinearity, and in this way 
these intrinsically localized modes may provide an excitation energy 
storage even at room temperatures where the  environment is strongly 
fluctuating.

Several theoretical models have been proposed in the study of the 
nonlinear dynamics and statistical mechanics of DNA (see the very 
comprehensive review \cite{grpd}). A particularly fruitful model was 
proposed by Peyrard and Bishop \cite{pb89} and Techera, Daemen and 
Prohofsky \cite{tdp89}. In the framework of this model the DNA molecule 
is considered to consist of two chains that are transversely coupled. 
Each chain models one of the two polynucleotide strands of the DNA molecule. 
A base is considered to be a rigid body connected with its opposite partner 
through the hydrogen-bond potential $V(u_n)$, where $u_n$ is the 
stretching of the bond connecting the bases,  $n$, and $n=0,\pm1,\pm2,..$ 
is labelling the base-pairs. The stretching of the $n'$th base-pair is 
coupled with the stretching of the $m'$th base-pair through a dispersive 
potential $U(u_n,u_m)$. The process of DNA denaturation was studied 
\cite{pb89,dpb93} under the assumption that the coupling between neighboring 
base-pairs is harmonic $U(u_n,u_{n+1})$. An entropy-driven denaturation 
was investigated \cite{dpbi93} taking into account a nonlinear potential 
$U(u_n,u_{n+1})$ between neighboring base-pairs.  The Morse potential was 
chosen \cite{pb89,dpb93,dpbi93} as the on-site potential $V(u_n)$, but 
also the longitudinal wave propagation and the denaturation of DNA has 
been investigated \cite{mu90} using the Lennard-Jones potential to describe 
the hydrogen bonds.

In the main part of the previous studies the dispersive interaction $U$
was assumed to be short-ranged and a nearest-neighbor approximation was 
used. It is worth noticing, however, that one of two hydrogen bonds
which is responsible for the interbase coupling, namely the hydrogen bond in 
the
$N-H...O$  group, is characterized by a finite dipole moment.  Therefore, 
a stretching of the base-pair will cause a change of the dipole moment,
so that the excitation transfer in the molecule will be due to transition 
dipole-dipole interaction with a $1/r^3$ dependence on the distance, $r$. 
It is also well known that nucleotides in DNA are connected by 
hydrogen-bonded water filaments \cite{dr71,cc81}. In this case an 
effective long-range excitation transfer may occur due to the 
nucleotide-water coupling. 

In the last few years the importance of the effect of long-range 
interactions (LRI) on the properties of nonlinear excitations was 
demonstrated in several different areas of physics 
\cite{bkz90,WKK93,VER94,aekm93,gmcr96,gfnm95}. Quite recently 
\cite{ga97}, we proposed a new nonlocal discrete NLS model with a power 
dependence on the distance of the matrix element of the dispersive interaction.
 It was found that there is an interval of bistability in the NLS models with
 a long-range dispersive interaction. One of these states is a continuum-like
soliton and the other is an intrinsically localized state. 


Another model, which we consider in this paper, describes an excitation energy 
(or excess electron) transport in a highly anharmonic molecular chain. 
This line of investigation is rather new, since until recently most 
attention has been concentrated on strongly elastic systems, such as the 
excitons in the $\alpha$-helix proteins. The subsonic Davydov solitons can 
exist in this case, and the harmonic approximation is enough for a 
consideration of the lattice deformation caused by the exciton 
\cite{Dav87,Revis}. But if the lattice is not supposed 
to be rigid or if the motion of the excitation at supersonic velocities 
is studied, the deformation cannot be considered small and 
anharmonic terms must be taken into account. For example, the anharmonic
 interactions play an important 
role in the dynamics of $DNA$. It is possible that they might concentrate 
vibrational energy in $DNA$ into soliton-like objects \cite{mu90,DNA2}.

The anharmonic Davydov model of the molecular chain 
has already been much studied during the last fifteen years 
\cite{Dav87,DZ83,DZ84,CEEG92,GCM95,ZSS95,ZSS96,Min97}. It is well known 
that there exist solitons of Davydov type and of Boussinesq type in
the system. P{\'e}rez and Theodorakopoulos \cite{PT87} 
investigated numerically an interaction between these solitons in 
the $\alpha$-helix proteins, and concluded that it inhibits the stability of 
the Davydov {\em vibrational soliton} at room temperature. 
But the effective anharmonicity parameter in that system is rather
small. The recent investigations on the subject \cite{GCM95,ZSS95,ZSS96,Min97}
demonstrated that in highly anharmonic chains (e.g. in the problem of
excess electron transfer in $\alpha$-helix proteins), the interaction
between Davydov and Boussinesq solitons changes its sign, and the
solitons can form a bound state. 


The outline of the paper is the following. 
In Sec.\ II we investigate the effects of long-range interactions on the 
nonlinear dynamics of the two-strand model of DNA. First, we reduce
the equations of motion of the model to the nonlocal discrete nonlinear 
Schr{\"o}dinger (NLS) equation. Then, we present the analytical theory 
and the results of numerical simulations of the discrete NLS model with a 
long-range dispersive interaction. We obtain stationary localized 
solutions, demonstrate the existence of the bistability phenomenon, 
and discuss their stability. The investigation of the switching between 
bistable states completes the section. We show that a  
controlled switching between narrow, pinned states and broad, mobile states 
with only small radiative losses is possible when the stationary states 
possess an internal breathing mode. 
In Sec.\ III we investigate the anharmonic Davydov model of the molecular 
chain and show that it reduces to the H\'enon - Heiles system \cite{Astr}, 
which is completely integrable at some fixed value of anharmonicity. 
There are three types of stationary solitons in this case: Boussinesq, 
Davydov, and a new type, namely a supersonic two-bell shaped soliton. 
We calculate the energies of the solitons. Then, we 
develop a variational approach for the non-integrable case.  
We show that in the case of highly anharmonic chains, the two-bell shaped 
soliton is an oscillating bound state of Davydov and Boussinesq solitons. 
This bound state is caused by the excitation (or electron) tunnelling in 
the effective two-well potential which is created by the exciton 
(electron) -- phonon interaction and anharmonic terms in the lattice 
potential. On the contrary, at small anharmonicity Davydov and Boussinesq 
solitons repel each other, and the bound state cannot exist. Finally, we
test the dynamical stability of the bound state at various values of 
anharmonicity numerically. 

\section{Effects of long-range dispersion: 
bistability of localized excitations}

\subsection{System and equations of motion}

In this section we study the two-strand model of DNA which 
is described by the Lagrangian 
\begin{eqnarray}
\label{1} 
L=T-U_{BP}-U_{LR} \; ,
\end{eqnarray}
where
\begin{equation}
T=\frac{M}{2}\,\sum_n \left(\frac{d u_n}{d t}\right)^2
\label{2}
\end{equation}
is the kinetic energy and $M$ is the mass of the base-pair, 
\begin{equation}
\label{4}
U_{BP} = \sum_n V(u_n)
\end{equation}
is the potential energy which describes an intrabase-pair interaction,
and 
\begin{equation} 
\label{3} 
U_{LR} = \frac{1}{4} \, \sum \!\!\!\!\!\! \sum_{n, m (n \neq m)} \!\!\! 
J_{n,m}(u_m -u_n)^2 
\end{equation}
is the long-range dispersive interbase-pair interaction of the
stretchings. 
In Eqs.\ (\ref{2})--(\ref{3}) $n$ and $m$ are base-pair indices. 
The value $u_n=0$ of the base-pair stretching corresponds to the 
minimum of the intrabase-pair potential $V(u_n)$. 
We investigate the model with the following power dependence
on the distance (assuming that the lattice constant equals unity) 
of the matrix element of the base elastic coupling 
\begin{equation}
\label{5}
J_{n,m}=\frac{J}{|n-m|^s} \; , 
\end{equation}
where the constant $J$ characterizes the strength of the coupling, and 
$s$ is a parameter being introduced to cover different physical situations 
including the nearest-neighbor approximation ($s=\infty$), 
quadrupole-quadrupole ($s=5$) and dipole-dipole ($s=3$) interactions. 
In the case of the DNA molecule the long-range interaction is 
evolved from the existence of charged groups in the DNA molecule. 
To take into account a possibility of screening of the interaction 
or an indirect coupling between base-pairs (e.g. via water filaments), 
we shall consider also the case when the matrix element of the base 
elastic coupling has the Kac-Baker \cite{Baker61,KH73} form 
\begin{equation}
\label{6}
J_{n,m}=J\,e^{-\beta |n-m|} \; ,
\end{equation} 
where $\beta$ is the inverse radius of the interaction. 

Assuming that 
\begin{eqnarray}
\label{7}
\left.
\frac{\partial^2 V(u_n)}{\partial u_n^2} \right|_{u_n=0}\,
\gg\, \left. 
\frac{\partial^j V(u_n)}{\partial u_n^j} \right|_{u_n=0}
\end{eqnarray}
for $j=3, 4, ...$, 
that is the anharmonicity of the intrabase-pair potential is rather small, 
we shall use a rotating-wave approximation 
\begin{equation}
\label{8}
u_n=\psi_n\,e^{-i \Omega\,t}+c.c. \; ,
\end{equation}
where $\left. \Omega=\sqrt{V''(u_n)/M} \right|_{u_n=0}$ is the frequency of 
the harmonic oscillations and $\psi_n(t)$ is the complex amplitude which 
is supposed to vary slowly with time. Inserting Eq.\ (\ref{8}) into 
Eqs.\ (\ref{1})--(\ref{3}) and averaging with respect to the fast 
oscillations of the frequency $\Omega$, we conclude that the effective 
Lagrangian of the system can be represented in the form 
\begin{eqnarray}
\label{lagra}
{\cal L}=\frac{i}{2} \sum_n\left(\dot \psi_n\psi^*_n-\dot 
\psi_n^*\psi_n \right)- {\cal H} \; ,
\end{eqnarray}
where the dot denotes the differentiation with respect to the 
rescaled time $\tau=t/(2 M \Omega)$.
Here 
\begin{eqnarray}
\label{9} 
{\cal H}={\cal U}_{BP}+{\cal U}_{LR} \; 
\end{eqnarray}
is the effective Hamiltonian of the system, where 
\begin{equation} 
\label{10} 
{\cal U}_{LR} = \frac{1}{2} \, \sum \!\!\!\!\!\! \sum_{n, m (n \neq m)} 
\!\!\! J_{n,m}|\psi_m -\psi_n|^2 
\end{equation}
is the effective dispersive energy and
\begin{eqnarray}
{\cal U}_{BP} &=& \sum_n \Biggl( \frac{\Omega}{2\pi} 
\int_{0}^{2\pi/\Omega} V(\psi_n\,e^{-i\Omega\,t}+c.c) dt 
\nonumber \\ &-& M \Omega^2\,|\psi_n|^2 \Biggr) 
\label{11}
\end{eqnarray}
is the effective intrabase-pair potential. Usually either a Morse 
potential \cite{pb89,dpb93} or a Lennard-Jones potential \cite{mu90} 
is used to model the hydrogen bonds. With these potentials however it 
is very complicated to obtain any analytical results. Therefore, to gain 
insight into the problem we shall use a simplified nonlinear potential 
in the form
\begin{equation}
\label{17}
{\cal U}_{BP} =-\frac{1}{(\sigma +1)} \sum_n 
|\psi_n|^{2(\sigma +1)} \; , 
\end{equation}
where the degree of nonlinearity $\sigma$ is a parameter which we 
include to have the possibility to tune the nonlinearity as well. 

From the Hamiltonian (\ref{9}) we obtain the equation of motion 
$i \dot {\psi_{n}}= \ds{\frac{\partial {\cal H}}{\partial \psi_{n}^*}}$ 
for the wave function $\psi_{n}(\tau)$ in the form 
\begin{eqnarray}
\label{12} 
i \dot{\psi_{n}}+ \sum_{m (m \neq n)} J_{n,m} (\psi_{m} - \psi_{n})+ 
|\psi_n|^{2\sigma} \psi_n=0 \; .
\end{eqnarray}
The Hamiltonian ${\cal H}$ and the number of excitations 
\begin{equation}
\label{13}
N=\sum_n |\psi_n|^2
\end{equation}
are conserved quantities. 

Thus we have reduced the equations of motion
of the initial model of DNA dynamics to the nonlocal discrete nonlinear
Schr{\"o}dinger (NLS) equation studied quite recently in detail
\cite{ga97,mj97}. In what follows we shall review the obtained results 
in application to the DNA molecule. 

Looking for the stationary solutions of Eq.\ (\ref{12}) of the form 
\begin{equation}
\label{sta}
\psi_n=\phi_n \exp (i \Lambda \tau)
\end{equation}
with a real shape function $\phi_n$ and frequency $\Lambda$, we
obtain the equation for $\phi_n$ in the form 
\begin{equation}
\label{14}
\Lambda \phi_n= \sum_{m (m \neq n)} J_{n,m} (\phi_m-\phi_n)+
\phi_n^{(2\sigma +1)} 
\end{equation}
with $J_{n,m}$ described by Eqs.\ (\ref{5}) or (\ref{6}). 
Thus Eq.\ (\ref{14}) is the Euler-Lagrange equation for 
the problem of extremizing ${\cal H}$ under the constraint $N=constant$. 

\subsection{Bistability of stationary states}

To develop a variational approach to the problem we use an ansatz for 
a localized state in the form 
\begin{equation}
\label{15}
\phi_n=\sqrt{N\tanh\alpha} \exp(-\alpha |n|) \; ,
\end{equation}
where $\alpha$ is a trial parameter. This ansatz is chosen to satisfy 
automatically the normalization condition (\ref{13}), so that the 
problem of extremizing ${\cal H}$ under the constraint $N=constant$ 
is reduced to the problem of solving the equation 
$\ds{\frac{d{\cal H}}{d\alpha}=0}$.

Inserting the trial function (\ref{15}) into the Hamiltonian given by
Eqs.\ (\ref{9}), (\ref{10}), (\ref{17}) and (\ref{5}) and evaluating 
the discrete sums which enter in these equations (see \cite{ga97} for 
details), we get the dispersive part of the Hamiltonian 
\begin{eqnarray}
{\cal U}_{LR} &=& 2 N J \{\zeta(s)- \tanh(\alpha) F(e^{-\alpha}, s-1)
\nonumber \\ &-& F(e^{-\alpha}, s) \} \; 
\label{19}
\end{eqnarray}
and  the intrabase-pair potential 
\begin{eqnarray}
{\cal U}_{BP} &=& -\frac{N^{\sigma+1}}{\sigma+1} \,f_{\sigma} \; , 
\nonumber \\ \quad \mbox{with} \quad 
f_{\sigma}&=&\tanh^{\sigma+1} (\alpha) \,\coth [(\sigma +1) 
\alpha ] \; , 
\label{20}
\end{eqnarray}
where 
\begin{equation}
\label{22}
\zeta(s) = \sum_{n=1}^{\infty} n^{-s} \quad \mbox{and} \quad 
F(z, s)=\sum_{n=1}^{\infty}(z^n/n^s)
\end{equation}
are Riemann's zeta function and Jonqi\`{e}re's function, 
respectively. 

According to the variational principle we should satisfy the condition 
$\displaystyle{\frac{d {\cal H}}{d \alpha}=0}$ which yields 
\begin{eqnarray}
\label{23}
N^{\sigma}=2 (\sigma+1) J \, (\tanh (\alpha) &F(e^{-\alpha},s-2)&
\\ \nonumber 
+ \tanh^2(\alpha) &F(e^{-\alpha},s-1)& )
\left( \frac{d f_{\sigma}}{d\alpha} \right)^{-1} \; .
\end{eqnarray}
As a direct consequence of Eq.\ (\ref{14}), the frequency $\Lambda$ 
can be expressed as 
\begin{eqnarray}
\label{24}
\Lambda=-\frac{1}{N}\, ({\cal U}_{LR}+2 \, {\cal U}_{BP})
\end{eqnarray}
with ${\cal U}_{LR}$ and ${\cal U}_{BP}$ being defined by 
Eqs.\ (\ref{19}) and (\ref{20}). 

\begin{figure}
\centerline{\hbox{
\psfig{figure=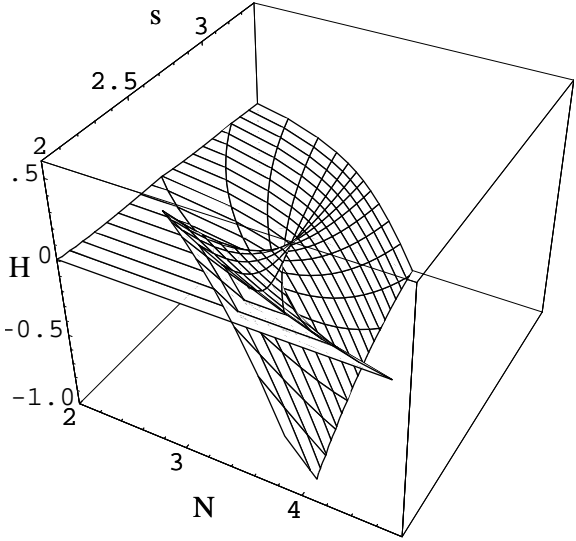,height=170mm,width=120mm,angle=0}}}
\vspace{-45mm}
\caption{The dependence of the energy $H$ of stationary soliton 
solutions versus the excitation number $N$ and the dispersion 
parameter $s$, obtained from the variational approach for 
$\sigma=1$.}
\label{fig:lr:h-n-s}
\end{figure}

The results of the above developed variational approach are plotted in
Figure \ref{fig:lr:h-n-s}, which shows the energy of the localized excitations 
as a function of $N$ and $s$ for the particular case $\sigma=1$. 

\vspace{-45mm}

One can see that there is a critical value $s_{cr}=2.72$ of the
dispersion parameter $s$ which separates two qualitatively
different regions. The region $s>s_{cr}$ is characterized by the usual
monotonic behavior of the soliton energy vs number of excitations
$N$. Thus the main features of all discrete NLS models with dispersive 
interaction $J_{n,m}$ decreasing faster than $|n-m|^{-s_{cr}}$
coincide qualitatively with the features obtained in the nearest-neighbor
approximation where only one stationary state exists for any $N$. 
But for $2<s<s_{cr}$ the bifurcation ``swallow tail'' takes place
introducing a multistability into the system. One can see that in this
case there is an energy interval where for each value of ${\cal H}$ 
there exist three stationary states with different excitation numbers.
The direct numerical solution of Eq.\ (\ref{14}) validates this
conclusion and gives the precise critical value $s_{cr}=3.03$. 
It is noteworthy that similar results are obtained for the Kac-Baker 
dispersive interaction (\ref{6}). In this case the multistability takes 
place for $\beta<1.70$. 

\vspace{2mm}
\begin{figure}
\centerline{\hbox{
\psfig{figure=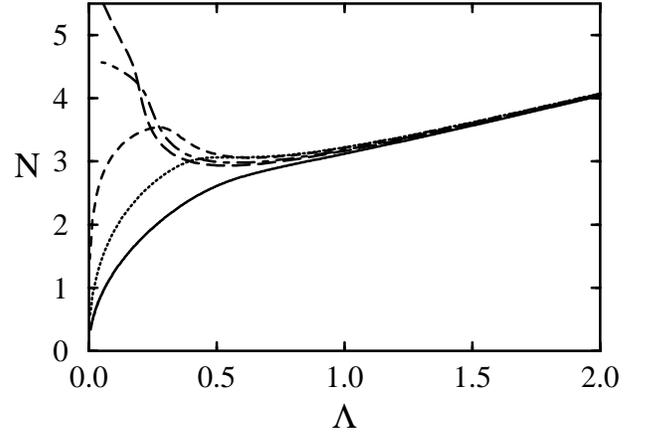,height=60mm,width=80mm,angle=0}}}
\caption{Number of excitations, $N$, versus frequency, $\Lambda$, 
numerically from Eq.\ (\ref{14}) for $\sigma=1$ and $s=\infty$ (full), 
3 (dotted), 2.3 (dashed), 2 (short-long-dashed), and 
1.9 (long-dashed).}
\label{fig:lr:norm-lambda}
\end{figure}
\vspace{2mm}
\begin{figure}
\centerline{\hbox{
\psfig{figure=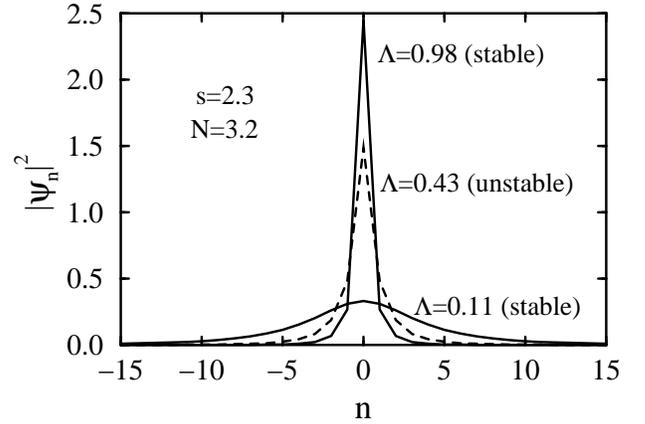,height=60mm,width=80mm,angle=0}}}
\caption{Shapes of the three stationary states for $s=2.3$ and $N=3.2$.
The stable: $\Lambda=0.11$ and $\Lambda=0.98$ (full). 
The unstable: $\Lambda=0.43$ (dashed).}
\label{fig:lr:forms}
\end{figure}

Figure \ref{fig:lr:norm-lambda} shows the dependence $N(\Lambda)$ obtained 
numerically for $\sigma=1$. This dependence demonstrates the same
transition from the monotonic behavior for $s>s_{cr}$ to the 
non-monotonic (${\cal N}$-type) behavior for $s<s_{cr}$ and assists to 
determine the stability of the stationary states. According to the 
theorem which was recently proven \cite{lst94}, the necessary and 
sufficient linear stability criterion for the stationary states is 
\begin{equation}
\label{25}
\frac{dN}{d\Lambda}=\frac{d}{d \Lambda}\, \sum_n\, \phi_n^2 > 0 \; . 
\end{equation}

Therefore we can conclude that only two stationary states are stable.
The third state, with intermediate value of $\Lambda$ and maximal
value of energy, is unstable. Thus it is reasonably to talk about
bistability of the system. 

The observed bistability is very similar to the one recently observed 
 \cite{lst94,Malomed}, where the nearest-neighbor case with an 
arbitrary degree of nonlinearity $\sigma$ was studied. The bistability 
appears in this case for $\sigma$ above a certain critical value. 

Figure \ref{fig:lr:forms} shows that the shapes of these solutions 
differ significantly. The low-frequency states are wide and continuum-like, 
 while the high-frequency solutions represent intrinsically localized 
states with a width of a few lattice spacings. 

Now we turn to discuss stationary states of the discrete NLS model given
by Eq.\ (\ref{14}) with arbitrary degree of nonlinearity. The main 
properties of the system remain unchanged, but the critical value $s_{cr}$ 
of the dispersion parameter is now a function of $\sigma$. The 
results of analytical consideration confirmed by simulation show that 
$s_{cr}$ increases with increasing $\sigma$. In particular, for 
$\sigma \geq 1.4$ (the value at which the discrete symmetrical ground state 
can be unstable in the nearest-neighbor approximation \cite{lst94}) 
the bistability in the nonlinear energy spectrum occurs even for $s\leq 6$.

\subsection{Switching between bistable states}

Having established the existence of bistable stationary states in the 
nonlocal discrete NLS system, a natural question that arises concerns
the role of these states in the full dynamics of the model. In particular,
it is of interest to investigate the possibility of switching between
the stable states under the influence of external perturbations, and what
type of perturbations that could be used to control the switching.
Switching of this type is important in the description of nonlinear 
transport and storage of energy in biomolecules like the DNA, since a 
mobile continuum-like excitation can provide action at distance, while 
the switching to a discrete, pinned state can facilitate the structural 
changes of the DNA \cite{geor96,grpd}. 

\begin{figure}
\centerline{\hbox{
\psfig{figure=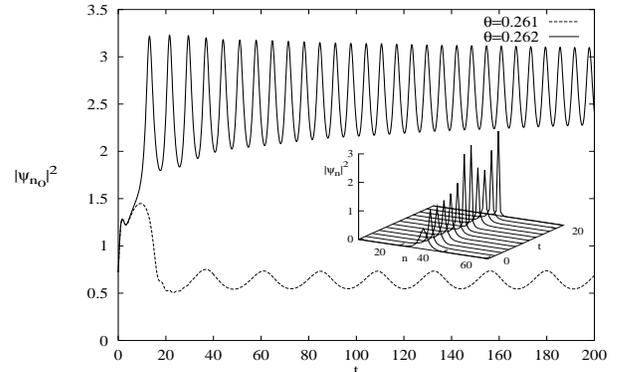,height=50mm,width=80mm,angle=270}}}
\caption{Switching from continuum-like to discrete state for the NLS model 
with the Kac-Baker long-range interactions (\ref{6}) where $\beta=1$. 
The initial state $\phi_n$ has the frequency $\Lambda\simeq 0.310$ and
$N=3.6$. We plot a time evolution of $|\psi_{n_0}(\tau)|^2$ when a phase 
torsion is applied to the center site with $\theta=0.261$ (lower curve) and 
$\theta=0.262$ (upper curve), respectively; inset shows time 
evolution of $|\psi_{n}(\tau)|^2$ for $\theta=0.262$.}
\label{fig:lr:switch1}
\end{figure}

\begin{figure}
\centerline{\hbox{
\psfig{figure=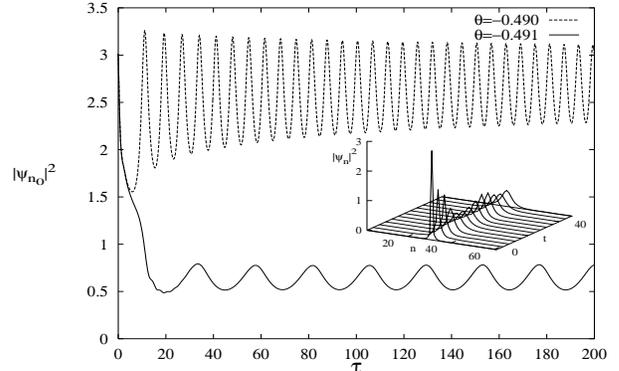,height=50mm,width=80mm,angle=270}}}
\caption{Switching from discrete to continuum-like state for the NLS model 
with the Kac-Baker long-range interactions (\ref{6}) where $\beta=1$. 
The initial state $\phi_n$ has the frequency $\Lambda\simeq 1.423$ and
$N=3.6$. We plot a time evolution of $|\psi_{n_0}(\tau)|^2$ when a phase 
torsion is applied to the center site with $\theta=-0.490$ 
(upper curve) and 
$\theta=-0.491$ (lower curve), respectively; inset shows time 
evolution of $|\psi_{n}(\tau)|^2$ for $\theta=-0.491$.}
\label{fig:lr:switch2}
\end{figure}

An illustration of how the presence of an internal breathing mode can 
affect the dynamics of a slightly perturbed stable stationary state is 
given in Figures \ref{fig:lr:switch1} and \ref{fig:lr:switch2}. 
To excite the breathing mode we 
apply a spatially symmetrical, localized perturbation, which we choose to 
conserve the number of excitations in order not to change the effective 
nonlinearity of the system. The simplest choice, which we have used in the 
simulations shown here, is to kick the central site $n_0$ of the system at
$t=0$ by adding a parametric force term of the form 
$\theta\delta_{n,n_0}\delta(\tau)\psi_n(\tau)$ to the left-hand-side 
of Eq.\ (\ref{12}). All details and a physical motivation of the appearance 
of such kind of parametric kick are described in Ref.\ \cite{mj97}. 
As can be easily shown, this perturbation affects only the site $n_0$ 
at $\tau=0$, and results in a 'twist' of the stationary state at this 
state with an angle $\theta$, i.e. $\psi_{n_0}(0)=\phi_{n_0}\,e^{i\theta}$. 
The immediate consequence of this kick is, as can been deduced from the 
form of Eq.\ (\ref{12}), that $\ds{\frac{d}{d\tau}}\left(|\psi_{n_0}|^2\right)$ 
will be positive (negative) when $\theta>0$ ( $\theta<0$). Thus, 
to obtain switching from the continuum-like state to the discrete state we 
choose $\theta>0$, while we choose  $\theta<0$ when investigating switching
in the opposite direction. We find that in a large part of the 
multistability regime there is a well-defined threshold value $\theta_{th}$, 
such that when the initial phase torsion is smaller than $\theta_{th}$, 
periodic, slowly decaying 'breather' oscillations around the initial state
will occur, while for strong enough kicks (phase torsions larger than
$\theta_{th}$) the state switches into the other stable stationary state.

\vspace{2mm}

\vspace{2mm}

It is worth remarking that the particular choice of perturbation is not
important for the qualitative features of the switching, as long as  there 
is a substantial overlap between the perturbation and the internal breathing
mode. We also believe that the mechanism for switching described here
can be applied for any multistable system where the instability is 
connected with a breathing mode. For example, we observed \cite{mjo97} 
a similar switching behavior in the nearest neighbor discrete NLS equation 
with a higher degree of nonlinearity $\sigma$, which is known 
\cite{lst94} to exhibit multistability. 

\section{Effects of anharmonicity: 
bound state of Davydov and Boussinesq solitons}

\subsection{System and equations of motion}

Let us consider in this section an excitation energy (or excess 
electron) transport in the anharmonic molecular chain. 
In Davydov's coherent states approximation \cite{Dav87} the 
Hamiltonian of the system takes on the form 

\begin{eqnarray}
\label{anh:hamil}
H & = & -J \sum_n \left[ \psi_n^* ( \psi_{n+1}+\psi_{n-1} )-2
        | \psi_n |^2 \right] \nonumber \\
  &   & +\chi \sum_n
        | \psi_n |^2 (\beta_{n+1} - \beta_{n-1}) \nonumber \\
  &   & + \frac{1}{2} \sum_n \Bigl[ M \left( \frac{d \beta_n}{dt} 
  \right)^2 + \omega (\beta_{n+1} - \beta_n )^2 \nonumber \\
  &   & - \frac{2}{3} \alpha 
  \omega (\beta_{n+1} - \beta_n)^3 \Bigr] \; .
\end{eqnarray}
Here, $\psi_n(t)$ is the excitation wave function of the molecule at site
$n$, $\beta_n(t)$ is the displacement of the $n$-th molecule from
equilibrium, $J$ is the matrix element of the excitation transition,
the constant $\chi$ characterizes the exciton - displacement
interaction, $M$ is the mass of the molecule, and the parameters $\omega$
and $\alpha$ characterize the elasticity and the anharmonicity of the
lattice, respectively.

The equations of motion for $\psi_n(t)$ and $\beta_n(t)$ are
\begin{eqnarray}
i \hbar \frac{d}{dt} \psi_n(t) & = & -J(\psi_{n+1}+\psi_{n-1}-2\psi_n)
\nonumber \\
                               &   & + \chi (\beta_{n+1}
                                     - \beta_{n-1})\psi_n \; , 
                                     \nonumber \\
M \frac{d^2}{dt^2} \beta_n(t)  & = & \omega ( \beta_{n+1}+\beta_{n-1}
                                     - 2 \beta_n ) 
\nonumber \\
                               &   & \times (1-\alpha (
                                     \beta_{n+1} - \beta_{n-1} ) )
                                       \nonumber \\
                               & + & \chi ( |\psi_{n+1}|^2
                                     - |\psi_{n-1}|^2 ) \; .
\end{eqnarray}
The Hamiltonian (\ref{anh:hamil}) conserves the number of
excitations in the chain.  We shall assume that there is only one
excitation. Thus the normalization condition for the wave function
$\psi_n(t)$ is
\begin{equation}
\label{anh:norm}
\sum_n | \psi_n(t) |^2 =1 \; .
\end{equation}
Using the continuum limit
$\psi_n(t) \to e^{ikx} \psi(x,t)$ and
$\beta_n(t) \to \beta(x,t)$, where $x=n \ell$ and $\ell$ is the
lattice constant, we shall consider the solutions in the form of 
travelling waves
\begin{eqnarray}
&\psi (x,t) =  (8 \Delta)^{-1/2} \;
e^{i \Lambda t} \; \varphi(\vartheta) \; , & \nonumber \\
&\partial \beta / \partial x = - (J \cos(k \ell) / \chi) \; 
u(\vartheta) \; , & 
\end{eqnarray}
where $\Delta=3 \chi^2/(\omega J \cos ( k \ell ))$ and 
$\vartheta = (x-v t)/ \ell$. 
As a result we obtain the system of equations
\begin{eqnarray}
&\ddot{\varphi}-A \varphi + 2 u \varphi = 0 \; ,&
\label{anh:eqs-cont-1} \\
&\ds{\frac{d^2}{d \vartheta^2}}( \ddot{u} - 4Bu + g u^2 -
\varphi^2 ) = 0  \; ,&
\label{anh:eqs-cont-2}
\end{eqnarray}
where the parameters $A$ , $B$ and $g$ are given by the expressions 
\begin{eqnarray}
A &=& ( 2 J + \hbar \Lambda ) / ( J \cos ( k \ell ) ) -  2 \; ,
\nonumber \\
B &=& 3 ( v^2 / v_0^2-1 ) \; , 
\nonumber \\ 
g &=& (12 \alpha J \cos ( k \ell ) ) / \chi \; .
\end{eqnarray}
Here $v_0=\ell \sqrt{\omega / M}$ is the sound velocity,
$v=(2J \ell / \hbar ) \sin (k \ell )$ is the soliton velocity,
and the dots denote the differentiation with respect to $\vartheta$ .
We shall assume that the effective mass of excitation is positive,
i.e.  $ J>0$,  and consider the carrier wave vector $k$ in
the interval $0 \leq k \ell < \pi/2$.

At the boundary conditions
\begin{center}
$ u( \pm \infty )=\dot{u}( \pm \infty )=\ddot{u}( \pm \infty )=
\varphi( \pm \infty )=\dot{\varphi}( \pm \infty )=0 \; , $
\end{center}
the equations of motion (\ref{anh:eqs-cont-1}) and 
(\ref{anh:eqs-cont-2}) are similar to that of the H\'enon-Heiles 
system \cite{Astr} except for the sign in front to $\varphi^2$ in 
Eq.\ (\ref{anh:eqs-cont-2}). This similarity has been used in our
investigations \cite{CEEG92,GCM95,Min97} of the system --- we shall 
review the results in what follows. An alternative approach to 
the problem has been developed recently by Zolotaryuk, Spatschek, 
and Savin \cite{ZSS95,ZSS96}.

\subsection{The soliton solutions in the completely 
integrable case}

It is well known that in a general case the H\'enon-Heiles system 
\cite{Astr} is not completely integrable. However in the following 
three cases \cite{Liht}:
\begin{quote}
   (i)   $A = 4B$ , $g=1$ , \\
   (ii)  $4A = B$ , $g=16$ , \\
   (iii) $g=6$ and arbitrary $A$ and $B$ , 
\end{quote}
there exists a second integral of motion, and thus it is a
Liouville completely integrable system.
The conditions (i) and (ii) can be satisfied for only one value of the
soliton velocity, since the normalization condition (\ref{anh:norm}) 
imposes a link between parameters $A$ and $B$. But we intend to 
investigate how the soliton energy and shape depend on the soliton 
velocity. Therefore we shall consider here only the third case. 
It was shown in Refs.\ \cite{CEEG92,GCM95} that in this case there exist only
three types of soliton solutions of Eqs.\ (\ref{anh:eqs-cont-1}) and 
(\ref{anh:eqs-cont-2}), namely:

\medskip

(a) The Boussinesq soliton:
\begin{equation}
\label{anh:sol-Bq}
\varphi = 0 \; \; , \; \; \; \; \; \;
u = \frac{6}{g} B \, \sech^2 \sqrt{B}\vartheta \; . 
\end{equation}
It exists at supersonic velocities for any $g$, and
represents a lattice compression which moves along the chain
without changing its shape.

\medskip

(b) The Davydov soliton \cite{DZ83}:
\begin{equation}
\label{anh:sol-i}
\varphi = 2 \sqrt{A(A-B)} \, \sech \sqrt{A} \vartheta \; \; , \; \; \;
\; \; u = A \, \sech^2 \sqrt{A} \vartheta \; . 
\end{equation}
It exists at subsonic ($B<0$) as well as at supersonic ($B>0$) 
velocities, but only for $g=6$. 

\medskip

(c) The two-bell shaped soliton, which 
exists only at supersonic ($B>0$) velocities and $g=6$:
\begin{eqnarray}
\varphi &=& 2 \sqrt{A} (A-B) S^{-1}( \vartheta , R ) \cosh \sqrt{B} (
\vartheta - R ) \; , \nonumber \\
u &=& \frac{d^2}{d \vartheta^2} \ln S(\vartheta , R ) \; ,
\label{anh:sol-ii}
\end{eqnarray}
where $R$ is an integration constant and
\begin{eqnarray}
\label{anh:S}
S ( \vartheta , R ) &=& \sqrt{A} \cosh ( \sqrt{B} ( \vartheta - R ))
\cosh ( \sqrt{A} \vartheta ) \nonumber \\
&-& \sqrt{B} \sinh ( \sqrt{B} ( \vartheta - R )) \sinh ( \sqrt{A}
\vartheta ) \; .
\end{eqnarray}
This solution was found analytically and studied numerically 
in Refs.\ \cite{CEEG92,GCM95}, 
where it was interpreted as a bound state of Davydov and Boussinesq 
solitons. It was also found numerically in Refs.\ \cite{ZSS95,ZSS96} 
where the authors believe, in contrast to our conclusion, that it is 
a bound state of a Davydov soliton with {\em two} Boussinesq solitons. 

\begin{figure}
\centerline{\hbox{
\psfig{figure=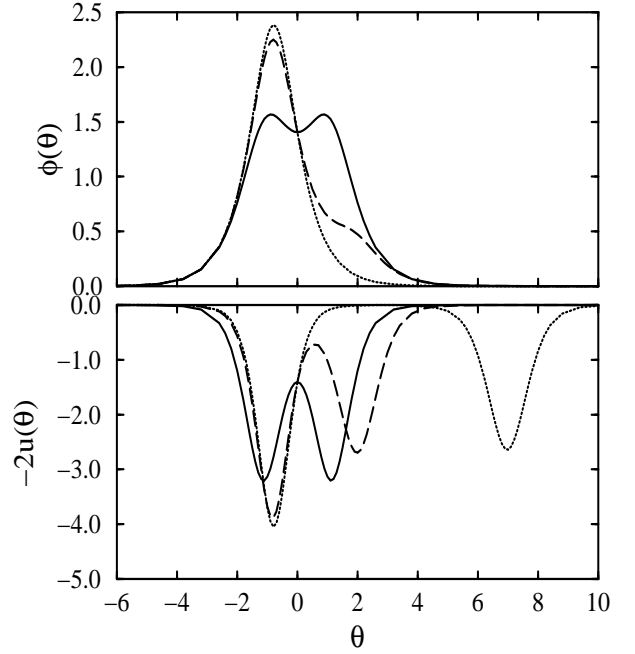,height=90mm,width=80mm,angle=0}}}
\caption{The form of the two-bell shaped solitons at $\Delta=1$ and 
$v=1.2 v_0$ for different values of distance $R$: $R=0$ (solid), 
$R=1$ (dashed), and $R=6$ (dotted).}
\label{fig:anh:form}
\end{figure}

It is interesting to note that the deformation function (\ref{anh:S})
has the form of the two-soliton solution of the KdV--equation
\cite{Lamb}. So we can conclude that the excitation $\varphi (
\vartheta )$ is similar to that of a quantum particle moving in
two-well potential $-2u(\vartheta)$ (see Figure \ref{fig:anh:form}). 
One of the wells is created by the lattice soliton, and the second
is caused by the interaction of the excitation with the lattice. 
Part of the time the particle lives in the well that was dug by itself, 
then it tunnels to the well that was created by the lattice soliton, 
and so on. As a result of such a complicated behavior we obtain a wave 
function $\varphi ( \vartheta ) \; $ in the form of Eq.\ 
(\ref{anh:sol-ii}) (see Figure \ref{fig:anh:form}).
It is interesting to remark that at the velocity within the range 
\begin{equation}
v_0 < v < v_0 [ 1 + ( \Delta^2 / 54 )^{1/3} ]^{1/2} \; ,
\end{equation}
the wave function of the soliton has the one-bell shape at any distance
$R$ between the wells in the potential $-2u( \vartheta )$. On the 
contrary, if the velocity exceeds $v_0 [ 1+ ( \Delta/2)^{2/3} ]^{1/2}$ 
we obtain the two-bell shape at any $R$. However, the ratio of the soliton 
maximal values in this case is proportional to 
$\sech \sqrt{A} R$ at $|R| \gg 1 / \sqrt{B}$. 
Thus we may conclude that the excitation (or electron) lives mostly
in the well that was dug out by itself, tunnelling to the other well 
with an exponentially small rate.

\vspace{2mm}

\begin{figure}
\centerline{\hbox{
\psfig{figure=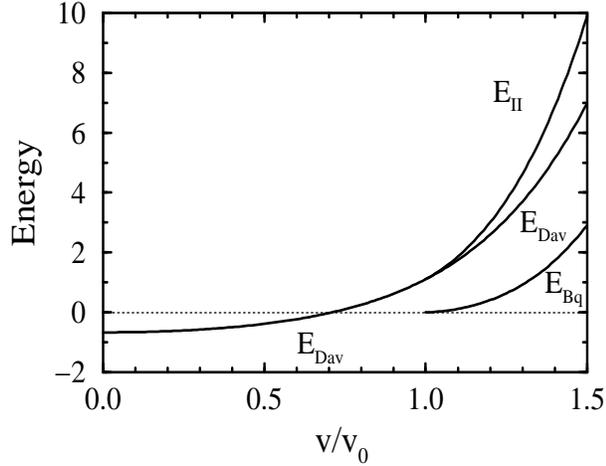,height=65mm,width=80mm,angle=0}}}
\caption{The energy of the Davydov ($E_{Dav}$), Boussinesq ($E_{Bq}$) 
and two-bell shaped ($E_{II}$) solitons vs the velocity, for 
$\Delta=10$.}
\label{fig:anh:energy}
\end{figure}

We have calculated (see Figure \ref{fig:anh:energy}) the velocity 
dependence of the soliton energy for all three types of 
solitons at $g=6$. For the energy of Boussinesq soliton 
(\ref{anh:sol-Bq}) we get 
\begin{equation}
E_{Bq} = \frac{2 \omega}{3} \left( 1+ \frac{4}{15} B 
\right) B^{3/2} \; .
\end{equation}
The energy of Davydov soliton (\ref{anh:sol-i}) is given by the 
expression
\begin{equation}
E_{Dav} = E_{ex} + J \cos ( k \ell ) \left[ \frac{2}{5 \Delta} 
\left(1+4 \frac{v^2}{v_0^2} \right) A^{3/2} - \frac{3}{10} A 
\right] \; ,
\end{equation}
where
\begin{equation}
E_{ex} = \frac{\hbar^2 v^2}{8J \ell^2} \cos ( k \ell ) \;
\end{equation}
is the exciton energy and $A$ is determined by the equation
\begin{equation}
\sqrt{A} (A-B) = \Delta \; .
\end{equation}
Finally the energy of the two-bell shaped soliton turned out to be
an exact sum of the energies of the Davydov and Boussinesq solitons: 
\begin{equation}
E_{II} = E_{Dav} + E_{Bq} \; .
\end{equation}
It is interesting that this energy does not depend on the distance 
$R$. It means that at $g=6$ the Davydov and Boussinesq solitons do
not interact. However, as it will be shown below they 
interact in the non-integrable case $g \neq 6$. Namely, they repel
each other at $g<6$ and form a bound state at $g>6$. 

\vspace{2mm}

\subsection{Variational approach in the non-integrable case}

In what follows we shall develop a variational approach to the
investigation of the behavior of the two-bell shaped solitons in
the anharmonic chain far from the completely integrable case.
Considering the Hamiltonian (\ref{anh:hamil}), one can see that 
the Lagrangian of the chain has the form
\begin{eqnarray}
& L &  =
       \sum_n \frac{i \hbar}{2} [ \psi_n^* \partial_t \psi_n 
       - \psi_n \partial_t \psi_n^* ]
       \nonumber \\
&   &  + \sum_n \psi_n^* [J ( \psi_{n+1} + \psi_{n-1} - 2 \psi_n ) 
\nonumber \\
&   &  - \chi (
       \beta_{n+1} - \beta_{n-1} ) \psi_n ] \nonumber \\
&   &  + \frac{M}{2}
       \sum_n ( \partial_t \beta_n )^2
\nonumber \\
&   &  - \frac{\omega}{2} \sum_n \left[ ( \beta_{n+1} - \beta_n )^2 -
       \frac{2 \alpha}{3} ( \beta_{n+1} - \beta_n )^3 \right] \; .
\end{eqnarray}
Substituting into it the solution (\ref{anh:sol-ii}) of Eqs.\ 
(\ref{anh:eqs-cont-1}) and (\ref{anh:eqs-cont-2}) as a trial 
function ($A$ and $B$ are supposed to be constant and $R$ is time
dependent), we obtain an effective Lagrangian $L(R , \partial_t R )$. 
If the soliton velocity is close to the sound
velocity ($B \ll A$) the energy of the two-bell shaped soliton is
\begin{eqnarray}
\label{anh:ener-var}
E_{II} \simeq E_{Dav} + E_{Bq} + J \cos ( k \ell ) 
         \frac{B^{3/2}}{\Delta} \left(\frac{d R}{d \tau} \right)^2 
         \nonumber \\
      + J \cos ( k \ell ) \frac{g-6}{18} \left[ \frac{4}{5} 
            \Delta^{2/3} + B \{ 4- \sech^2 ( \sqrt{B} R ) \} \right] 
            \; . 
\end{eqnarray}
The equation of motion in this case is 
\begin{equation}
\frac{d^2 R}{d \tau^2} + \frac{\nu^2}{\sqrt{B}}
\frac{\sinh ( \sqrt{B} R )}{\cosh^3
( \sqrt{B} R )} = 0 \; ,
\end{equation}
where
$\nu^2 =(g-6) \sqrt{B} \Delta / 18$
and $\tau = \sqrt{\omega / M} \, t$ is the dimensionless time.

One can see that at $g<6$ there is an intersoliton repulsion. The 
distance $R \to \infty$ when $\tau \to \infty$, and the 
two-bell shaped soliton is dissociated into Davydov and Boussinesq 
solitons.

On the contrary, at $g>6$ there is a potential well in Eq.\ 
(\ref{anh:ener-var}), and the distance $R$ between solitons is a 
periodic function 
\begin{equation}
\label{anh:periodic}
\sinh ( \sqrt{B} R ) =
\sinh ( \sqrt{B} R_0 ) \sin \frac{2 \pi}{T} ( \tau - \tau_0 ) \; ,
\end{equation}
with the period of oscillations 
\begin{equation}
T = \frac{2 \pi}{\nu}
\cosh ( \sqrt{B} R_0 ) \; .
\end{equation}
Here $R_0$ is the maximum value of $R$ and $\tau_0$ is the initial time.

\vspace{2mm}
\begin{figure}
\centerline{\hbox{
\psfig{figure=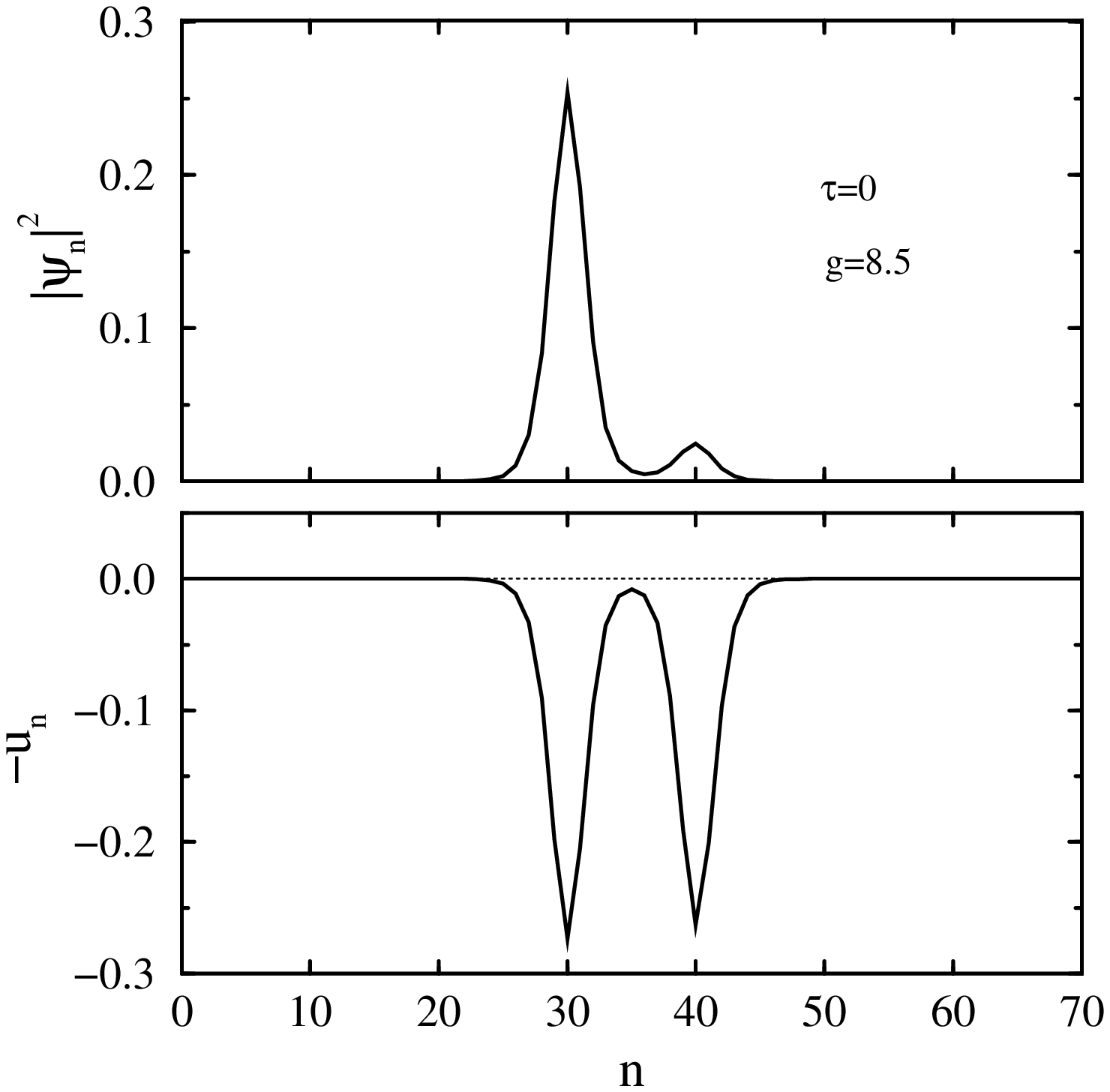,height=95mm,width=80mm,angle=0}}}
\centerline{\hbox{
\psfig{figure=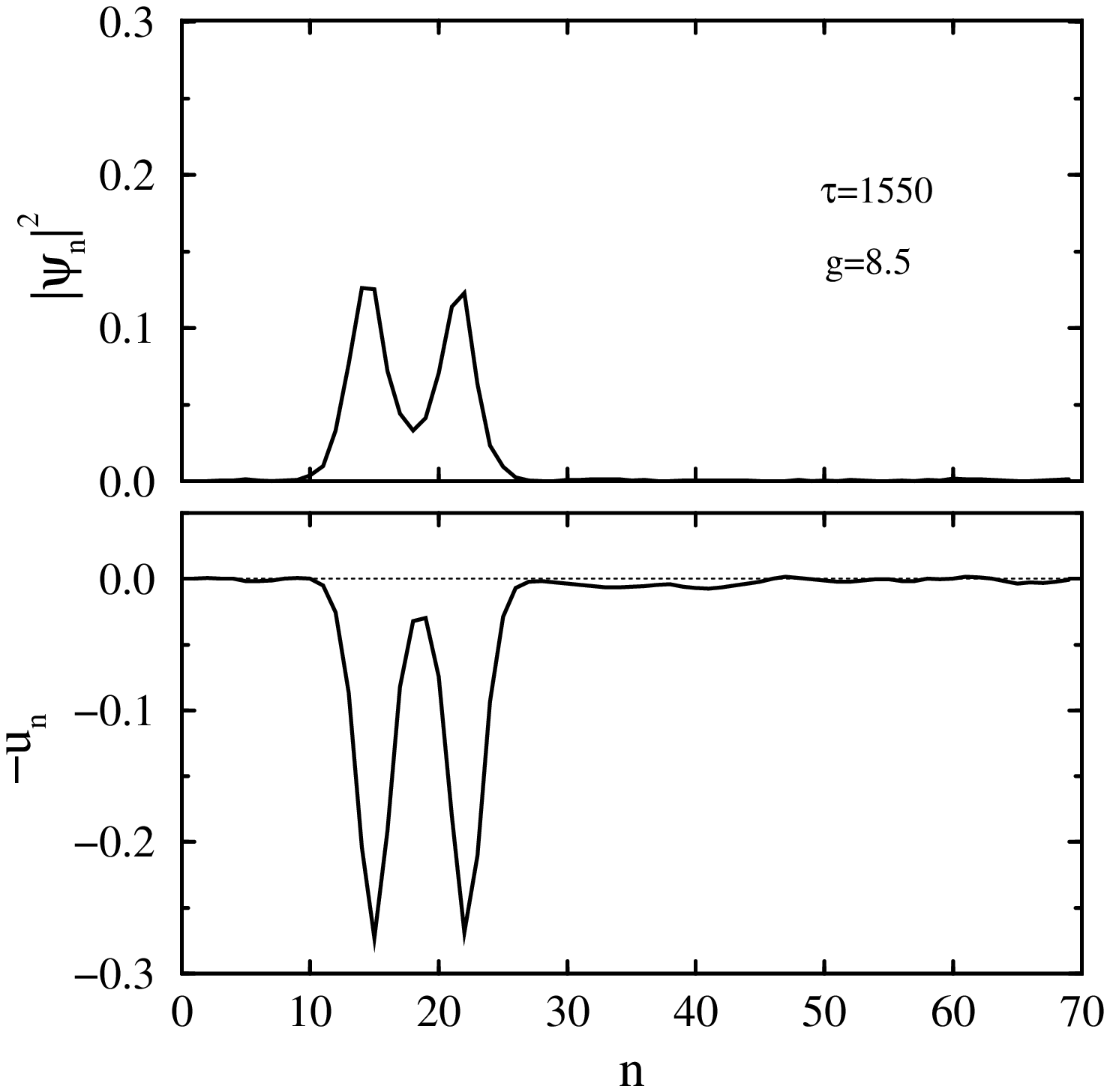,height=95mm,width=80mm,angle=0}}}
\centerline{\hbox{
\psfig{figure=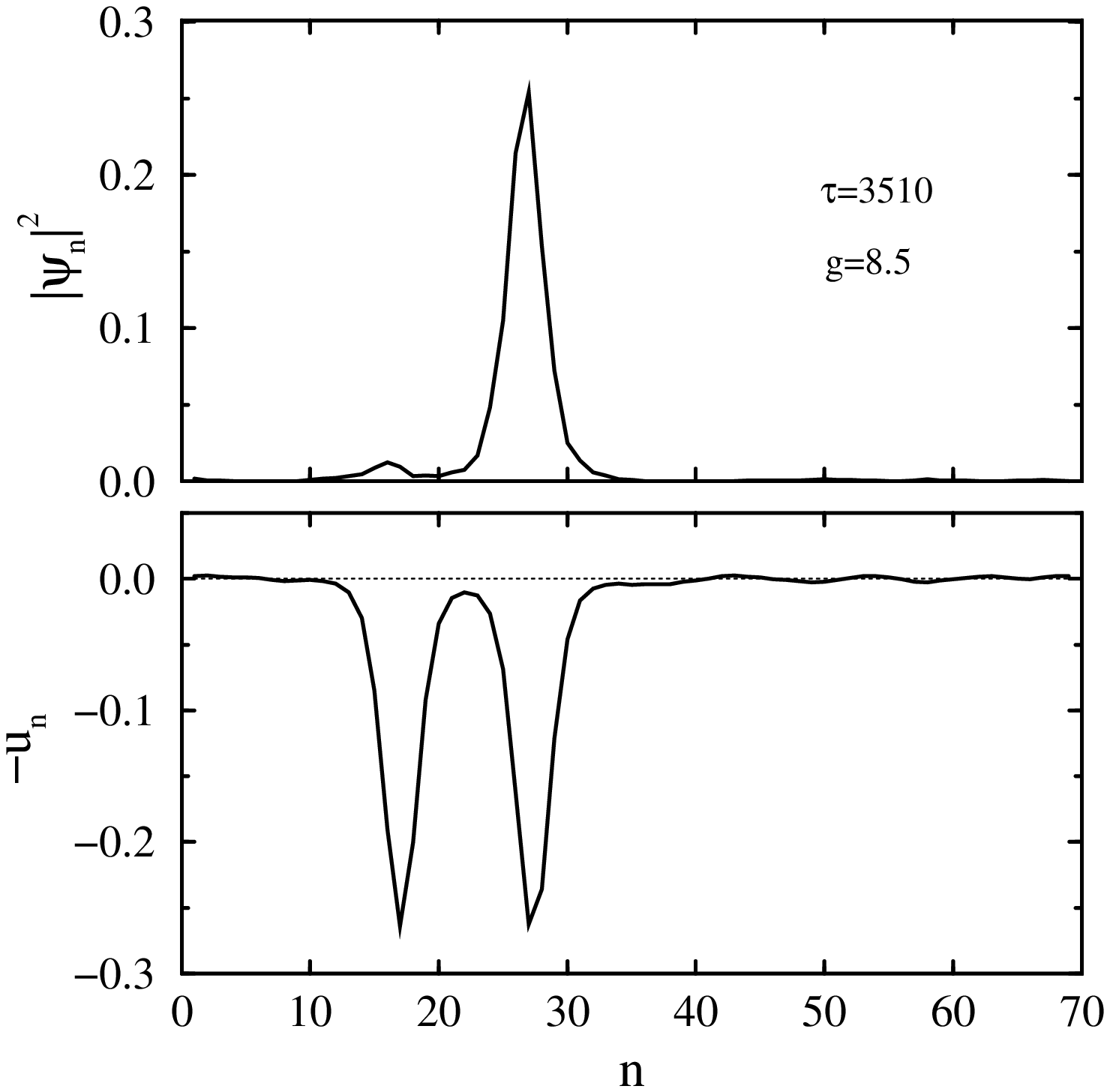,height=95mm,width=80mm,angle=0}}}
\centerline{\hbox{
\psfig{figure=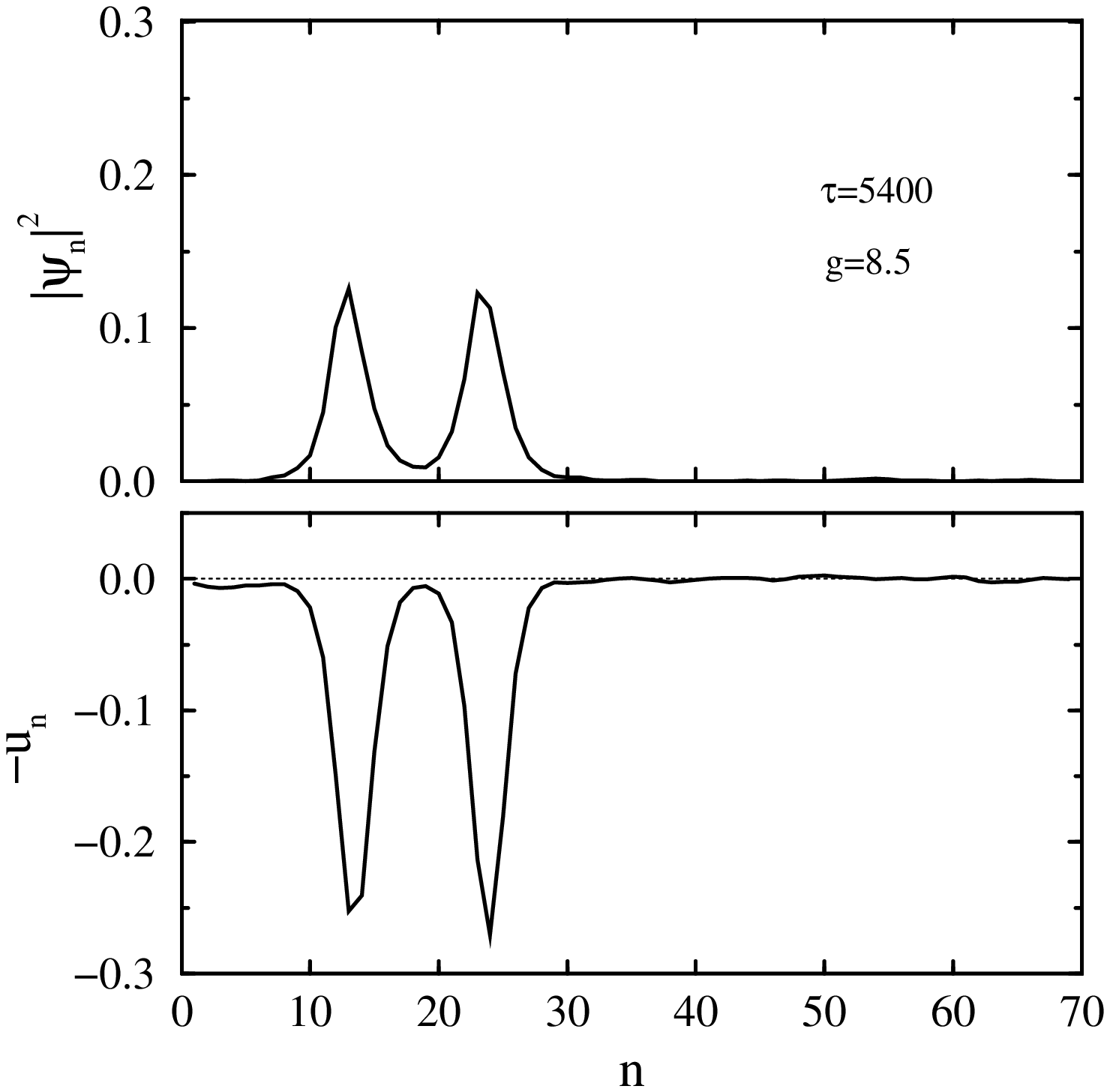,height=95mm,width=80mm,angle=0}}}
\centerline{\hbox{
\psfig{figure=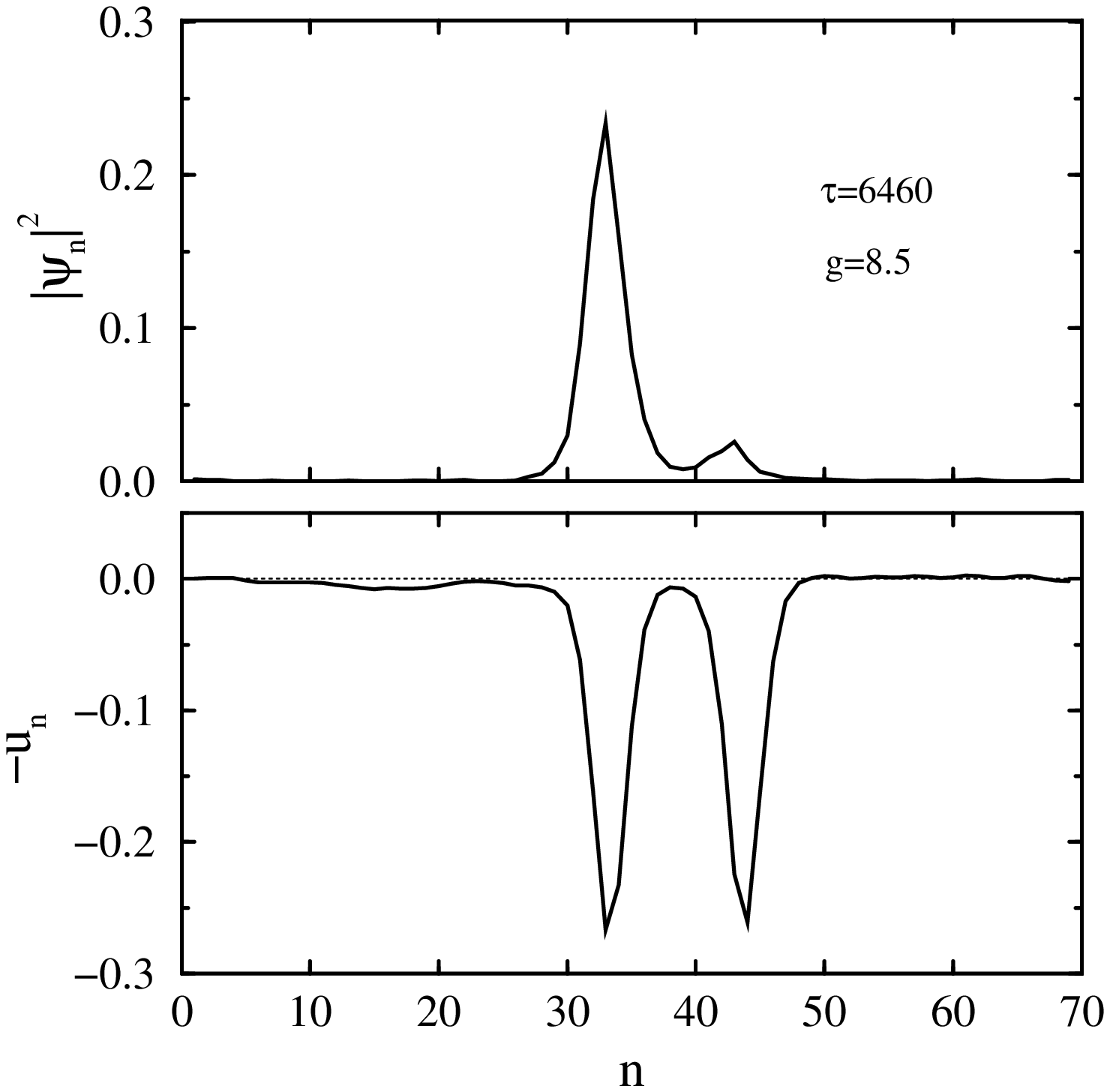,height=95mm,width=80mm,angle=0}}}
\caption{Numerical calculations of the dynamics of two-bell shaped 
solitons for $g=8.5$ at different moments of dimensionless time: 
(a) $\tau=0$, (b) $\tau=1550$, (c) $\tau=3510$, (d) $\tau=5400$, 
(e) $\tau=6460$. The boundary conditions are periodic.}
\label{fig:anh:big-g}
\end{figure}

Thus, we can conclude that in highly anharmonic molecular chains, there 
should exist a bound state of Davydov and Boussinesq solitons. 
The dynamical stability of this bound state has been tested numerically. 
In Figure \ref{fig:anh:big-g} we present the results of numerical
simulations for $g=8.5$. As an initial state we chose an asymmetric 
two-bell shaped soliton. One can see that as time increases, the bells 
approach each other and their heights (as well as the depths of wells) 
equalize. Thereafter, the distance between the bells increases, and a mirror 
reflected asymmetric two-bell shaped soliton appears. Thus, the Davydov 
and Boussinesq solitons form a rather stable bound state and demonstrate an 
oscillating motion in accordance with Eq.\ (\ref{anh:periodic}). 

In conclusion, we note that for the Davydov vibrational soliton motion 
in the $\alpha$--helix molecule the parameter $g \simeq 0.8$, so that 
the bound state of supersonic Davydov and Boussinesq 
soliton cannot exist. But for the electron motion in the polypeptide 
molecule we get $g \ge 12$. Thus we arrive at the conclusion that 
the bound state of the supersonic Davydov electrosoliton and the 
Boussinesq soliton can exist in the $\alpha$--helix biopolymers.

\section{Conclusion}

We have considered two models for energy and charge transport 
and storage in biomolecules. In these models we took into account the
long-range dispersive (first model) and anharmonic (second model) 
interactions, and showed that these interactions are responsible for the 
existence of new types of excitations. 

We have proposed a new nonlocal discrete nonlinear 
Schr{\"o}dinger (NLS) model for the modelling of the nonlinear dynamics 
of the DNA molecule with long-range ($r^{-s}$ and $e^{-\beta r}$) dispersive 
interaction between its charged groups. We have shown that when the
long-range interactions decay slowly, there is an energy interval where 
two stable stationary states exist at each value of the Hamiltonian 
${\cal H}$. One of these states is a continuum-like soliton and the 
other one is an intrinsically localized mode. By this means a bistability 
phenomenon came into existence in the system. This phenomenon is a result 
of the competition of two length scales: the 
usual scale of the NLS model, which is related to the competition between 
nonlinearity and dispersion (expressed in terms of the ratio $N/J$), and 
the radius of the long-range interactions. 

We have shown that a controlled switching between narrow (pinned) states
and broad (mobile) states is possible. Applying a perturbation in the form 
of a parametric kick, we demonstrated that switching occurs beyond some 
well-defined threshold value of the kick strength. The particular choice of 
perturbation is not important for the qualitative features of the switching, 
as long as  there is a substantial overlap between the perturbation and the 
internal breathing mode. Thus, we believe that the mechanism for switching 
described here can be applied for any multistable system where the 
instability is connected with a breathing mode. The switching  phenomenon 
could be important for controlling energy storage and transport in DNA 
molecules.

The second model is proposed for modelling of anharmonic biological 
macromolecules. We have shown that when the value of the effective 
(dimensionless) anharmonicity parameter is large enough, 
a bound state of Davydov and Boussinesq solitons can exist. 
For the Davydov {\em vibrational soliton} motion 
in the $\alpha$--helix proteins the anharmonicity parameter 
is too small, so that the bound state cannot exist. But for the {\em excess 
electron} motion in the polypeptide molecule it is rather large, and a 
bound state of the supersonic Davydov 
{\em electrosoliton} and the Boussinesq soliton can be formed. Furthermore, 
a slightly generalized version of the $DNA$ model \cite{mu90} which takes 
into account next neighbors transversal interactions between Toda chains 
actually yields strong anharmonicity. Hence there is a wide range of values 
for the anharmonicity parameter $g$ occurring in real systems.

\section*{Acknowledgments}

 Yu.G. and S.M. acknowledge support from the Ukrainian Fundamental 
Research Foundation (Grant No.~2.4/355). Yu.G. acknowledges also 
partial financial support from SRC QM "Vidhuk". M.J. acknowledges 
financial support from the Swedish Foundation STINT. 


\end{multicols}
\end{document}